\documentstyle[12pt]{article}
\def\newmathop#1{\mathop{#1}\limits}
\def\too#1{\displaystyle\newmathop{#1}_{q \to 1}}
\begin{document}
\title{
\begin{flushright}
{\small SFU-HEP-108-93 }\\
\end{flushright}
%\vspace{0.2cm}
Coherent States, Dynamics and \\
Semiclassical Limit on Quantum Groups}
\author{ I.Ya. Aref'eva \thanks{Permanent address:  Steklov Mathematical
Institute, Vavilov st.42, GSP-1,117966, Moscow , Russia; E-mail :
arefeva@qft.mian.su},
 R. Parthasarathy \thanks{ Permanent address:
 The Institute of Mathematical Sciences, Madras
 600 113, India, E-mail: sarathy@imsc.ernet.in}, \\ K.S. Viswanathan
 \thanks{E-mail: kviswana@sfu.ca}
 and I.V.Volovich\thanks{Permanent address:  Steklov Mathematical
Institute, Vavilov st.42, GSP-1,117966, Moscow , Russia; E-mail:
volovich@mph.mian.su}\\
Department of Physics, Simon Fraser University\\
Burnaby, British Columbia, V5A 1S6, Canada}
\date {$~$}
\maketitle
\begin{abstract}
   Coherent states on the quantum group $SU_q(2)$ are defined by using
harmonic analysis and representation theory of the algebra of functions
on the quantum group. Semiclassical limit $q\rightarrow 1$ is discussed
and the crucial role of special states on the quantum algebra in an
investigation of the semiclassical limit is emphasized. An approach to
$q$-deformation as a  $q$-Weyl quantization and a relavence of contact
geometry in this context is pointed out.
Dynamics on the quantum group parametrized by a real time variable  and
corresponding to classical rotations is considered.
\end{abstract}
\newpage

\section{Introduction}
   Recently attempts at constructing field theoretic models with a quantum
group playing the role of the gauge group have been made
\cite{AV}-\cite{AAr}. In
contrast to applications of quantum groups \cite{Dr}-\cite{Fad}
for solutions of standard
models in field theory and statistical physics   the point here is an
attempt to build a new class of field theory models still preserving the
standard Minkowski (or Riemannian) space-time. Such a development would
be a natural mathematical application of the idea of quantum groups. As a
possible physical motivation one notices a plausible mechanism of symmetry
breaking which could be an alternative to the Higgs mechanism \cite{AV}, a
derivation of the Weinberg angle in terms of the parameter $q$ \cite{IP1} and a
clarification of the special role of $U(2) = SU(2)\times U(1)$ symmetry
group \cite{AAr}. Quantum group chiral fields, i.e. fields taking values in a
quantum group have also been discussed \cite{AV,AB,AVI,Luk}. To give a
rigorous meaning to
such theories one still needs to clarify many questions even before
quantization (one distinguishes $q$-deformation which corresponds to
`classical' theories on `quantum' groups and `$\hbar $-deformations'
which involves quantization of such a theory, see \cite{AVII}). In
particular, one has to understand whether there exist nontrivial
functions from classical space-time to the quantum groups and what it
means for a variable to belong to a quantum group. See also a recent
consideration
of this issue by Frishman, Lukierski and Zakrzewski \cite{Luk}. Another
important issue is the question of dynamics on q- and $\hbar$-deformed
phase space \cite{AVII}-\cite{Manko}

   Recall that according to \cite{Dr}-\cite{Fad},\cite{Mad} a quantum group is
considered, as it should be, as a `noncommutative manifold' with a
coordinate ring which is a Hopf algebra with a given set of generators.
Therefore one thinks that there are only few points on a quantum group and
it is not at all clear how one gets from this object the usual Lie group
which is a smooth manifold. It is this question which we will
discuss in this paper.

   Our approach is the following. Consider the analogous question for the
case of a simple quantum mechanical problem, like the harmonic oscillator. In
quantum mechanics we also have an algebra with only a given set of fixed
generators, say operators of position $\hat{x}$ and momentum $\hat{p}$.
How can we get from these two fixed operators the classical phase space
with two real variables $x$ and $p$? The answer of course is well known.
To get correspondence with classical theory one should take the average of
the operators $\hat{x}$ and $\hat{p}$ with respect to appropriate states,
for example the coherent states. So the information about the classical
variables is encoded into appropriate states. The same approach we will
use for quantum groups. We define coherent states on quantum group
algebra $SU_q(2)$ depending on an element $u=(u_{mn})$ of the classical group
$SU(2)$ and show the correspondence in the sense that,
\begin{equation}
<g_{mn} \Psi(u)>_q ~\rightarrow ~ u_{mn},\label {I1}
\end{equation}
when $q\rightarrow 1$. Here $g_{mn}$ are generators of the
algebra of functions on the quantum group $SU_q(2)$, the operator
$\Psi(u)$ defines a coherent state and the brackets in (\ref{I1}) stands for
the Haar
functional on $SU_q(2)$. The operator $\Psi(u)$ is
\begin{equation}
\Psi(u)= \sum _{j} (2j+1) \mbox {tr} (W^{j~*} T^j(u)).\label {I2}
\end{equation}
Here $T^j(u)\ =\ (D^j_{nm}(u))$ is a unitary representation of $SU(2)$
of spin $j$, matrices $W^{j}=(W^{j}_{mn})$ have entries from the algebra
${\it {A}}$ (see below). One has an analogous formula like (\ref{I1})
for an arbitrary
polynomial $f(g)$ with respect to the generators $g_{ij}$,
\begin{equation}
<f(g)\Psi(u)>_q ~\rightarrow ~ f(u),\ \ \ \ \ q\rightarrow 1.\label {I3}
\end{equation}
 Coherent states in quantum mechanics are well known \cite{Kl,Coh}.
 q-deformed
coherent states were considered in \cite{PVI}-\cite{Sol}. Coherent states on
classical Lie
groups are defined as \cite{Coh},
\begin{equation}
 \mid u > = T(u)\mid \phi >,\label {I4}
\end{equation}
where $\mid \phi >$ is a vector in the space of the unitary
representations of $T(u)$. It seems that formula (\ref{I2}) gives a natural
generalization of coherent states (\ref{I4}). However there are two important
differences between our definitions (\ref{I1}) and (\ref{I2}) and the
formula (\ref{I4}). The
first one is that one deals here with with a quantum group and so we have
operators
$W^j$ in (\ref{I2}). The second one is that formula (\ref{I4}) defines the
vector
$\mid \phi>$ as a pure state. One can talk about coherent vectors (not
necessarily states) in this context. In (\ref{I4}) one has really a state
because the Haar functional as we will discuss below is nothing but the
statistical partition function \cite{LQGG}. So in this case one really deals
with
a coherent state which is defined by means of a density matrix. In this
interpretation the parameter $q$ is equal to,
\begin{equation}
q= \exp(-\beta ),
\end{equation}
where $\beta$ is the inverse temperature, $\beta = 1/T$. One has an
interpretation of the quantum group $SU_q(2)$ as a model of Bose-gas at a
 temperature $T=-1/\ln q$. Note here that this interpretation of
the deformation parameter $q$ as temperature is different from the
considerations of q-Bosons under non-zero temperature \cite{Zachos}.

   We are not discussing here  coherent states on
the q-deformed universal enveloping algebras
\cite{Bied,Mac},\cite{PVI}-\cite{Sol},\cite{Ell}-\cite{Demos}.
We hope to clarify a
relation of these with  (1) and (2) in a forthcoming publication.

\section{Representations of Pol$(SU_q(2))$}
  An element of the quantum group $SU_q(2)$  for $0<q<1$ is a
    $2\times2$ matrix
$g=(g_{mn})$ which has the following canonical form
    \begin{equation}
g = (g_{mn})=\left( \begin{array}{lr}
a&-qc^* \\
c&a^* \end{array}\right),\label{c1}
\end{equation}
and satisfies the unitarity conditions
\begin{equation}
gg^* = g^*g~=~I.\label{c2}
\end{equation}
Here $a$ and $c$ are elements of some algebra with involution. Equations
(\ref{c2}) are
equivalent \cite {AV} to the following known relations \cite {Dr,Wor,Fad}
for the elements $a,c,a^{*}$ and $c^{*}$ :
\begin{equation}
ac = q ca, ~~ ac^* = q c^*a,~~cc^* = c^*c \label{cc1}
\end{equation}
\begin{equation}
 aa^* + q^2 cc^* = 1 ,~~
a^*a + c^*c =1.\label{c2.1}
\end{equation}
The algebra ${\it A} = Pol(SU_q(2))$ of polynomial functions on the
quantum group $SU_q(2)$ is generated as a C-algebra by elements $a, c,
a^* $ and $c^*$ with the relations (\ref{cc1}) and (\ref{c2.1}).

  ${\it A}$ is a Hopf algebra with the standard coproduct,
\begin{eqnarray}
\Delta : {\it A} &\rightarrow & \it {A}\otimes \it {A} \nonumber \\
\Delta (g_{mn}) &=&\sum_{k} g_{mk}\otimes g_{kn}
\end{eqnarray}
with counit $\varepsilon  :  {\it A}\rightarrow C$, $\varepsilon
(g_{mn}) = \delta _{mn}$, involution * and antipode $S:{\it A}\rightarrow
\it{A}$, $Sg = g^*$. All representations of the algebra $\it{A}$  by
operators in a Hilbert space were classified \cite{Wor,LS} and
there are only
the following series of unitary inequivalent nontrivial representations
parametrized by a real parameter $\phi$ , $0\leq \phi < 2\pi$.

These are in the space ${\cal H}={\cal L}^2(R)$ which is considered as the Fock
space for the oscillator. One has a basis ${\{e_n \}}^{\infty}_{k=0}$ in
${\cal H}$ and operators $a = a_{\phi}$ and $c = c_{\phi}$ acting as,
\begin{equation}
ae_{n} = \sqrt{1-q^{2n}} e_{n-1},~
ae_{0}=0;~~
ce_n = e^{i\phi} q^n e_n.
\end{equation}
There is also a trivial representation : $a_{\phi}=e^{i\phi}$,
$c_{\phi}=0$.
By introducing the standard creation and annihilation operators $b,b^{*}$ in
${\cal L}^{2}(R)$ satisfying $[b,b^{*}]=1$ ,
one can rewrite the representation (11) in the form
	\begin{equation}
		a=\sqrt{\frac{1-q^{2(N+1)}}{N+1}}b,~~
		c=e^{i\phi}q^{N}.
		\label{r}
	\end{equation}
Here  $N$ is the number operator, $N=b^{*}b$.

Recall that it is the left representation of the Hopf
algebra ${\it A }= Pol(G)$ which corresponds to the (right) representation of
the group $G$. Let $T: G\rightarrow GL(V)$ be a representation of the
group $G$, on a vector space $V$,
\begin{equation}
T(uu')~ = ~T(u) T(u'),~ u,u' \in G .\label{r2}
\end{equation}
Choose a basis ${\{ {\xi}_i\}}$ in $V$ and suppose that the corresponding
matrix elements are $W_{ij} \in Pol(G)$. Then one can rewrite (\ref{r2}) as ,
\begin{equation}
\Delta (W_{ij})={\sum}_k W_{ik} \otimes W_{kj}.\label{r3}
\end{equation}
Hence we have a linear map $\rho : V\rightarrow Pol(G)\otimes V$ defined
by,
\begin{equation}
\rho(e_n) ~=~ \sum _m D_{nm} \otimes e_m,
\end{equation}
satisfying,
\begin{eqnarray}
(\Delta \circ id)\circ \rho &=& (id \circ \rho)\circ \rho \nonumber \\
(\varepsilon \otimes id)\circ \rho &=& id.
\end{eqnarray}
A linear space $V$ is called a left corepresentation for the
Hopf algebra  $\it A$ if there exists a linear map $\rho : V\rightarrow
{\it A}\otimes V$ satisfying (15).

\section{Harmonic analysis on $SU_{q}(2)$ and q-Weyl quantization. }

We present now results on the representation theory and harmonic analysis
(Fourier transform) on  $SU_{q}(2)$, see \cite{Wor},\cite{LS}-\cite{Ko}.
First recall that for the group
$SU(2)$ for every dimension $2j+1$, where spin $j=0,\frac{1}{2},1,\frac{3}{2},
...$ there is one irreducible unitary representation $T^{j}(u)$. On $SU(2)$
there
exists an invariant Haar measure $du$, i.e.
\begin{equation}
\int f(u)du =\int f(u'u)du =\int f(uu')du ;~\int du =1
	\label{h1}
\end{equation}
Matrix elements $D^{j}_{mn}(u)$ of $T^{j}(u)$ taken with respect to an
orthonormal basis,
\begin{equation}
D^{j}_{mn}(u)=(e_{m}, T^{j}e_{n})
	\label{h3}
\end{equation}
are orthogonal:
\begin{equation}
\int D^{j}_{mn}(u)D^{j'*}_{m'n'}(u)	du=\frac{1}{2j+1}\delta _{jj'}
\delta _{mm'}\delta _{nn,}
	\label{h4}
\end{equation}
and any function $f(g)$ on the group $SU(2)$ can be expanded into
Fourier series
\begin{equation}
	f(u)=\sum _{j\in N/2} (2j+1)\sum _{m,n=-j}^{j}
\tilde{f}^{j}_{mn}	 D^{j}_{mn}(u)
	\label{h5}
\end{equation}
where
$$\tilde{f}^{j}_{mn}	= \int f(v)D^{j*}_{mn}(v)dv $$
One can represent an elements $u$ of the group  $SU(2)$ in the form
\begin{equation}
	\left( \begin{array}{lr}
\alpha &-\gamma ^* \\
\gamma &\alpha ^* \end{array}\right),~~~
\alpha ^* \alpha + \gamma ^* \gamma ^ =1,
	\label{h6}
\end{equation}
or using the Euler angles
\begin{equation}
	\alpha =e^{-\frac{1}{2}i(\phi +\psi)}
	\cos \frac{1}{2}\theta ,~~\gamma =e^{-\frac{1}{2}i(\phi -\psi)}
	\sin \frac{1}{2} \theta ,
	\label{h6a}
\end{equation}
$ 0\leq \phi ,\psi \leq 4\pi,~ 0\leq \theta \leq \pi$ and the Haar
measure is $du=\frac{1}{2} \sin  \theta d\theta \frac{d\phi }{4\pi}
 \frac{d\psi}{4\pi}$.

The representation $T^{j}(u)$
 can be realized as a representation in the space of all homogeneous
 polynomials of degree $2j$ of two complex variables. If $f( z_{1},z_{2} )$
 is such polynomial, then the operator $T^j (u)$ acts as follows
 \begin{equation}
T(u)f(z_{1},z_{2})=f(\alpha z_{1}+\gamma z_{2},
-\gamma ^{*}z_{1}+\alpha ^{*}z_{2})
 	\label{h7}
 \end{equation}
 One can take the following basis in the spin $j$ representation:
$$ e_{m}=\frac{\sqrt{2j!}}{\sqrt{  (j+m)!(j-m)!}}z^{j-m}_{1}z^{j+m}_{2}, $$
where $m\in I_{j}=\{-j,-j+1,...,j\}$. Then one has
\begin{equation}
T^j(u)e_m= \sum _{n} D^{j}_{mn}(u)e_n
	\label{h8}
\end{equation}
and
\begin{equation}
 D^{j}_{mn}(u)=\sqrt{\frac{(j+m)!(j-m)!}{(j+n)!(j-n)!}}
 \sum _{\mu}\left( \begin{array}{l}
j+n\\
\mu\end{array}\right)
\left( \begin{array}{l}
j-n\\
j-\mu -m\end{array}\right)\cdot
	\label{h9}
\end{equation}
$$ (\alpha )^{j+n-\mu} (\alpha ^{*})^{j-\mu -m} (\gamma )^{\mu}(-\gamma
^{*})^{m-n+\mu}.$$

The representations of  $SU_{q}(2)$ are described in a similar way. We
will use the explicit construction of these representations by
Masuda, Mimachi, Nagakami, Noumi and Ueno \cite{Jap}. Let
$V^{(j)}$  be a $C-$linear space with the basis
\begin{equation}
	\zeta ^{(j)}_{m}=\left[ \begin{array}{l}
2j\\
j+m\end{array}\right]^{1/2}_{q^{2}}a^{j-m}c^{j+m}
	\label{h10}
\end{equation}
where $j\in N/2 , m \in I_{j}$, $a$ and $c$ are generators of the algebra
${\it A}$,  and $\left[ \begin{array}{l}
m\\
n\end{array}\right]_{q}$ are the Gauss q-binomial coefficients:
\begin{equation}
\left[ \begin{array}{l}
m\\n\end{array}\right]_{q}= \frac {(q;q)_{m}}{(q;q)_{n}(q;q)_{m-n}},~~
(p;q)_{m}=\prod
_{k=0}^{m-1}(1-pq^{k})	, ~ (p;q)_{0}=1.	\label{h11}
\end{equation}
$V^{j}$ is a left A-comodule, i.e. there are elements
$W^{j}_{mn}=W^{j}_{mn}(g)$ from
${\it A}$  such that
\begin{equation}
	\Delta (\zeta ^{(j)}_{m})= \sum _{n\in I_{j}} W^{j}_{mn}\otimes \zeta
	^{(j)}_{n},
	\label{h15}
\end{equation}
\begin{equation}
	\Delta (W^{(j)}_{mn})= \sum _{n\in I_{j}} W^{j}_{mn}
	\otimes W^{(j)}_{kn}.
	\label{h16}
\end{equation}
There exists an explicit representation for the elements $ W^{j}_{mn}=
 W^{j}_{mn}(g)$
in terms of the little q-Jacobi polynomials \cite{Jap} as follows:

 If $m+n \leq 0, m\geq n : ~~ W^{j}_{mn}= $
 \begin{equation}
 q^{(j+n)(j-m)}\left[ \begin{array}{l}
j+m\\
j-n\end{array}\right]^{1/2}_{q^{2}}
\left[ \begin{array}{l}
j-n\\
m-n\end{array}\right]^{1/2}_{q^{2}}a^{-m-n}
c^{m-n}P^{m-n,-m-n}_{j+n}(c^{*}c;q^2) ,
 	\label{h17}
 \end{equation}

 If $m+n \leq 0, n\geq m :~~ W^{j}_{mn}= $
 \begin{equation}
q^{(j+m)(m-n)}\left[ \begin{array}{l}
j-m\\
n-m\end{array}\right]^{1/2}_{q^{2}}
\left[ \begin{array}{l}
j+n\\
n-m\end{array}\right]^{1/2}_{q^{2}}a^{-m-n}
(c^{*})^{n-m}P^{n-m,-m-n}_{j+n}(c^{*}c;q^2),
 	\label{h18}
 \end{equation}

 If $m+n \geq 0, n\geq m :~~ W^{j}_{mn}=$
 \begin{equation}
 q^{(n-m)(n-j)}\left[ \begin{array}{l}
j-m\\
n-m\end{array}\right]^{1/2}_{q^{2}}
\left[ \begin{array}{l}
j+n\\
n-m\end{array}\right]^{1/2}_{q^{2}}
P^{n-m,m+n}_{j-n}(-c^{*}c;q^2) (-c^{*})^{n-m}(a^{*})^{m+n},
 	\label{h19}
 \end{equation}

 If $m+n \geq 0, m\geq n :~~ W^{j}_{mn}= $
 \begin{equation}
q^{(m-n)(m-j)}\left[ \begin{array}{l}
j+m\\
m-n\end{array}\right]^{1/2}_{q^{2}}
\left[ \begin{array}{l}
j-n\\
m-n\end{array}\right]^{1/2}_{q^{2}}P^{n-m,-m-n}_{j+n}(c^{*}c;q^2)c^{m-n}
(a^{*})^{n+m}.
 	\label{h20}
 \end{equation}
Here the little q-Jacobi polynomials are defined by
\begin{equation}
	P^{(\alpha ,\beta)}_{n}(z;q)=\sum
	^{\infty}_{r=0}\frac{(q^{-n};q)_{r}(q^{\alpha +\beta +n+1};q)_{r}}
	{(q;q)_{r}(q^{\alpha +1};q)_{r}}(qz)^{r}.
	\label{h21}
\end{equation}
There exists a unique linear functional $h: A\to C$ with $h(f^{*}f)\geq 0$
for all $f\in A$ and $h(1)=1$ which is invariant, i.e. it satisfies the
condition
\begin{equation}
	(h\otimes id) \circ \Delta =e  \circ h=(id \otimes h) \circ \Delta\; .
	\label{}
\end{equation}
The functional $h$ on $A$ is the quantum Haar functional. We will denote
it
\begin{equation}
	h(f)=<f>_{q}, ~~f\in {\it A}
	\label{i1}
\end{equation}
By using the representation (12) for $a$ and $c$ operators it is
defined as
\begin{equation}
	<f>_{q}=\frac{ \mbox {Tr} fe^{-\beta H}}{ \mbox {Tr} e^{-\beta  H}},
	\label{i2}
\end{equation}
where
\begin{equation}
	 \mbox {Tr} f=\sum _{n=0}^{\infty }\frac{1}{2\pi}\int _{0}
^{2\pi } <n|f|n>d\phi 	,\label{i3}
\end{equation}
\begin{equation}
H=2N,
	\label{ham}
\end{equation}
and $|n> $ are n-particle oscillator states, $N|n>=n|n>$.
Therefore the quantum Haar functional is the thermodynamic average with
the Hamiltonian (\ref{ham}) (this is noted in \cite{LQGG}). In particular
the partition function is
\begin{equation}
	Z= \mbox {Tr} e^{-2\beta N }=\sum _{n=0}^{\infty}  e^{-\beta
	2n}=\frac{1}{1-e^{-2\beta } }.
	\label{i4}
\end{equation}

The Hopf algebra ${\it A}=Pol (SU_{q}(2))$ has an orthogonal decomposition
\begin{equation}
	{\it A}=\oplus _{j\in N/2} W^{j}
	\label{i5}
\end{equation}
with respect to $<.>_{q}$, where $ W^{j}$ is spanned by matrix elements $ W^{j}
_{mn}$. The matrix elements $ W^{j}_{mn}$ satisfy the following
orthogonality relations
\begin{equation}
< W^{j}_{mn} (W^{j'}_{m'n'})^{*}>_{q}=\delta ^{jj'}\delta _{mm'}\delta _{nn'}
\frac{q^{-2n}}{[2j+1]_{q}},
	\label{i6}
\end{equation}
where $$[n]_{q}=\frac{q^{n}-q^{-n}}{q-q^{-1}}.$$
One can consider Fourier transformation on $SU_{q}(2)$:
	$${\cal F}: {\it A}=Pol(SU_{q}(2))\to Mat (C) $$
\begin{equation}
	\label{i7}
	{\cal F}(f)=(\tilde {f}^{(j)})_{j\in N/2 }, ~
	\tilde {f}^{(j)}_{mn}=<f W^{j~*}_{mn}>_{q}.
\end{equation}
The inversion formula is given by
\begin{equation}
	f=\sum _{j\in N/2}[2j+1]_{q}  \mbox {Tr} _{q}(\tilde {f}^{(j)}W^{j~}),
	\label{i8}
\end{equation}
where the q-trace $ \mbox {Tr} _{q} M $  of a $(2j+1)\times (2j+1)$
matrix $M$ is given by
\begin{equation}
		\label{ i9}
 \mbox {Tr} _{q} M = \sum _{k\in I_{j}}q^{2k}M_{kk}.
\end{equation}

Let us recall that the standard Weyl quantization is based on the Fourier
analysis. If one has a function $f(x,p)$
on the classical phase space admitting the Fourier representation
\begin{equation}
f(x,p)=\int \tilde {f}(\zeta , \eta )e^{-i(x\zeta +p\eta)}d\zeta  d\eta
	\label{i10}
\end{equation}
then one defines a corresponding Weyl operator as
\begin{equation}
\hat {f}(X,P)=\int \tilde {f}(\zeta , \eta )e^{-i(X\zeta +P\eta)}d\zeta  d\eta
,
	\label{i11}
\end{equation}
where $X$ and $P$ are the usual position and momentum operators.

Analogously for a function $f(u)$ on $SU(2)$ one has the Fourier
representation (\ref{h5}), then a corresponding q-Weyl operator is
\begin{equation}
\hat {f}=Wf=\sum _{jmn}(2j+1)\tilde{f}^{j}_{mn}W	^{j}_{mn},
	\label{h12}
\end{equation}
which gives an element of the algebra ${\it A}$. In particular
$$Wu=g.$$
 One can develop now a
q-analog of the theory of pseudodifferential operators and the
Wigner-Moyal approach \cite{Flato} to quantum group.

\section{Coherent States on $SU_q(2)$}

We define a coherent state operator on $SU_q(2)$ as
\begin{equation}
\Psi (u)=\sum_{j\in N/2} \sum_{m,n\in I_j} (2j+1)W^{j*}_{mn}
D^{j}_{mn}(u)=
\sum_{j\in N/2} (2j+1) \mbox {Tr} W^{j*}
D^{j}(u^{-1}).	\label{ c1}
\end{equation}
Here $W^{j}_{mn}$ are matrix elements of the representation of $SU_{q}(2)$
of spin $j$ (i.e. the corepresentation of $A=Pol (SU_{q}(2))$) see
(\ref{h17} )-(\ref{h20} )
and $D^{j}_{mn}(u)$ are the $D$-functions  (\ref{h9} ), $u\in SU(2)$.
Instead of $(2j+1)$ one can put another constant $\lambda _{j}$ depending
on $q$ such that $\lambda _{j} \to (2j+1)$ when $q\to 1$. We are
considering $\Psi (u)$
as an analog of the operator $\exp ib^{*} z$ creating the standard
coherent states from the Fock vacuum, where $b^{*}$
is the creation operator and $z$ is a complex number.

Now we consider the "classical" limit for the coherent states. Let us
prove that one has the following limiting formula
\begin{equation}
\lim _{q\to 1}<f(g)\Psi (u)>_{q} =f(u).
	\label{c2a}
\end{equation}
Here $f(g)$
is an element from $A=Pol (SU_{q}(2))$ , i.e. a polynomial in  the
generators  $a^{*},a, c^{*}$ and $c$ with coefficients independent of $q$.
 For definiteness we assume the following ordering of the
arguments of the polynomial $f(a^{*},a, c^{*},c)$.  In each monomial let us
put first from the left $a^{*}$ of some degree, then $a$ and after that $
c^{*}$
and $c$. One can use also the "normal" ordering.

Using the definition of $\Psi (u)$ (\ref{c1}) and the Fourier expansion
(\ref{i8}) for $f(g)$ one sees that the proof of (\ref{c2a})
is reduced to proving the relation
$$\lim _{q\to 1}<f(g) W_{mn}^{j}>_{q} =\int f(u)D^{j*}_{mn}(u)du.
$$
First prove that
\begin{equation}
\lim _{q\to 1}<(a^{*})^{k_{1}}(a)^{k_{2}}(c^{*})^{k_{3}}(c)^{k_{4}}>_{q}=
\int (\alpha ^{*})^{k_{1}}(\alpha )^{k_{2}}(\gamma ^{*})^{k_{3}}
(\gamma)^{k_{4}} du ,	\label{c3}
\end{equation}
where $k_{i}$ are natural integers. Let us note that the left and the
right hand sides of (\ref{c3}) vanish identically
if $k_{1}\neq k_{2}$ and $k_{3}\neq k_{4}$. So we need only to
 consider
 \begin{equation}
<(a^{*}a)^{k_{1}}(c^{*}c)^{k_{3}}>_{q}
 	\label{c4}
 \end{equation}
Using (\ref {c2.1}) this is equal to
 \begin{equation}
<(1-c^{*}c)^{k_{1}}(c^{*}c)^{k_{3}}>_{q} ,
 	\label{c5}
 \end{equation}
i.e. one needs to consider only $<(c^{*}c)^{k}>_{q}$. It is equal to
 \begin{equation}
<(c^{*}c)^{k}>_{q} =(1-q^{2})\sum _{n=0}^{\infty}q^{2(k+1)n}=\frac{q^{-k}}
{[k+1]_{q}} .
 	\label{c6}
 \end{equation}
For the corresponding classical expression, by using (\ref {h6a})
one has
\begin{equation}
	\int (\gamma ^{*} \gamma )^{k}du=\frac{1}{2} \int _{0}^{\pi}
	(\sin \frac{1}{2}\theta )^{2k}\sin \theta d\theta=\frac{1}{k+1} .
	\label{c7}
\end{equation}
Comparing (\ref {c6})  and (\ref {c7}) one gets (\ref {c3}).

Now let us prove that
\begin{equation}
\lim _{q\to 1}<(a^{*})^{k_{1}}(a)^{k_{2}}(c^{*})^{k_{3}}(c)^{k_{4}}
W^{j}_{mn}>_{q}=
	\label{c8}
\end{equation}
$$\int (\alpha ^{*})^{k_{1}}(\alpha )^{k_{2}}(\gamma ^{*})^{k_{3}}
(\gamma )^{k_{4}}
D^{j*}_{mn}(u)du .$$
Since for any monomial one has the limiting relation (\ref {c3}) it is enough
to prove the relation
\begin{equation}
\lim _{q\to 1}W^{j}_{mn}(u)=D^{j*}_{mn}(u),
	\label{c9}
\end{equation}
where in $W^{j}_{mn}(u)$ we mean the expression
(\ref {h17})-(\ref {h21})in which the
quantum generators $a^{*},a, c^{*}$ and $c$ are replaced by their
classical counterparts $ \alpha ^{*},\alpha, \gamma ^{*}$ and $\gamma$.

Let us discuss the case   $m+m'\geq 0$, $m\geq m'$. In this case
\begin{equation}
	W^{j}_{mm'}(u)=q^{(m-j)(m-m')}\left[ \begin{array}{l}
j+m\\
m-n\end{array}\right]^{1/2}_{q^{2}}
\left[ \begin{array}{l}
j-n\\
m-n\end{array}\right]^{1/2}_{q^{2}}
		\label{c10}
\end{equation}
$$(\alpha
^{*})^{m+m'}(\gamma)^{m-m'}P^{(m-m',m+m')}_{j-m'}(\gamma ^{*}\gamma :
q^{2}).
 $$
 As one can expect,  from explicit formulae  (\ref {h21})  it follows that
the little q-Jacobi
 polynomials go to Jacobi polynomials when $q\to 1$:
 \begin{equation}
\lim _{q\to 1} P^{s;t}_{n} (z;q)=P^{s;t}_{n} (1-2z)	/P^{s;t}_{n} (1),
 	\label{c11}
 \end{equation}
and $$P^{s;t}_{n} (1)=\frac{(s+1)n}{n!}.$$
Here the Jacobi polynomials are defined by the formula
\begin{equation}
P^{s;t}_{n} (z)=\frac{(s+1)n}{n!}F(-n,n+s+t+1;s+1;\frac{1-z}{2})
	\label{c12}
\end{equation}
$$=\frac{(-1)^{n}}{2^{n}n!}(1-z)^{-s}(1+z)^{-t}\frac{d^{n}}
	{dz^{n}}[(1-z)^{n+s}(1+z)^{n+t}],$$
where $F$ is the hypergeometric function. Therefore one has
\begin{equation}
\lim _{q\to 1}W^{j*}_{mm'}(u)=(\frac{(j+m)!(j-m')!}{ (j+m')!(j-m)!} )^{1/2}
\frac{1}{(m-m')!}	\label{c13}
\end{equation}
$$(-\gamma ^{*})^{m-m'}(\alpha ^{*})^{m+m'}
F(m-j,j+m+1,m-m'+1,\gamma ^{*}\gamma ). $$
 One can show that the expression (\ref {c13}) is equal to (\ref {h9}).
To this end one needs to use the equality
\begin{equation}
F(m-j,j+m+1,m-m'+1,\gamma ^{*}\gamma )=	\label{c14}
\end{equation}
$$\sum _{\mu}(1-\gamma ^{*}\gamma
)^{j-m -\mu}(-\gamma ^{*}\gamma )^{\mu}\frac{(j+m')!}{\mu ! (j+m-\mu)! }
...\frac{(j-m)!}{ (j-\mu+m)!}\frac{(m-m')!}{ (m-m'+\mu)! }$$

\section{Dynamics on Quantum Groups}
Dynamics on the classical group $SU(2)$ is reduced to Euler rotations.
For the classical matrix $u =\left( \begin{array}{lr}
\alpha &-\gamma ^* \\
\gamma &\alpha ^* \end{array}\right) \in SU(2)$,
one has three basic motions:
\begin{equation}
 \alpha(t) = exp(-it)\alpha ,~~\gamma(t) = \gamma \label{d2a}
\end{equation}
\begin{equation}
\alpha(t) = \alpha ,~~ \gamma(t) = exp(-it)\gamma \label{d2b}
\end{equation}
\begin{equation}
 \alpha(t) = \alpha \cos t + \gamma \sin t ,~~
      \gamma(t) = \gamma \cos t - \alpha \sin t
         \label{d2c}
\end{equation}
Now we consider the corresponding "quantum", i.e. q-deformed motions.
First note that if $U_t$ is an unitary operator in $L_2(R)$, then the
matrix
\begin{equation}
	g(t) = (g_{mn}(t))=\left( \begin{array}{lr}
	a(t) & -qc^{*}(t) \\
	c(t) & a^{*}(t) \end{array} \right)
	\label{d3}
\end{equation}
belongs to $SU_q(2)$. Here,
\begin{equation}
	a(t) = U_t ~a~ U_t^{*} ,~~ c(t) = U_t~ c~ U_t^{*}.
	\label{d4}
\end{equation}
Therefore any unitary group operator $U_t$ gives a group of automorphisms
of $SU_q(2)$. One can take,
\begin{equation}
	U_t = e^{itH},
	\label{d5}
\end{equation}
 where $H$ is an arbitrary self adjoint operator depending on $ a, a^{*},
 c$ and $c^{*}$. The quantum dynamics is given by,
 \begin{equation}
 	g_{mn}(t)~=~e^{itH} g_{mn} e^{-itH}.
 	\label{d6}
 \end{equation}
We want to find such a Hamiltonian that in the semiclassical limit
$q\rightarrow 1$ one could have,
\begin{equation}
	<g_{mn}(t)\Psi(u)>_q \rightarrow u_{mn}(t),
	\label{d7}
\end{equation}
where $u_{mn}(t)$ is a classical rotation. For the infinitesimal rotation
one has,
\begin{equation}
	\delta g_{mn} ~=~[iH,g_{mn}]\epsilon,
	\label{d8}
\end{equation}
and,
\begin{equation}
	<[iH,g_{mn}]\epsilon \Psi(u)>_q \rightarrow \delta u_{mn}.
	\label{d9}
\end{equation}
To the first classical motion (\ref{d2a}) corresponds  the Hamiltonian
$H=N$. Indeed, in this case
\begin{equation}
	a(t) = e^{itN} a e^{-itN} = e^{-it}a;~~c(t) = c.
	\label{d10}
\end{equation}
and in the classical limit $q\rightarrow 1 $ one gets,
\begin{equation}
	<a(t)\Psi (u)>_q ~=~ e^{-it}<a\Psi(u)>\too
	{\longrightarrow }
	\alpha(t) =
	e^{-it}\alpha ,\label{d11}
\end{equation}
$$	<c(t)\Psi(u)>\too
	{\longrightarrow }\gamma(t) = \gamma .$$
To the classical motion (\ref{d2b}) one has the corresponding `quantum'
dynamics,
\begin{equation}
	a_{\phi}(t) = a_{\phi} ,~~ c_{\phi}(t) = e^{-it} c_{\phi}.
	\label{d12}
\end{equation}
In the semiclassical limit $q\rightarrow 1$ one gets,
\begin{equation}
	<a_{\phi}(t)\Psi(u)>_q\rightarrow \alpha ,~~ <c_{\phi}(t)\Psi(u)>_q
	\rightarrow e^{-it} \gamma.
	\label{d13}
\end{equation}
Now let us discuss rotations (\ref{d2c}). Take the following Hamiltonian,
\begin{equation}
	H_{\phi} = \frac{1}{1-q} (ae^{-i\phi} + a^{*}e^{i\phi}).
	\label{d14}
\end{equation}
Note the occurrence of the singular factor $\frac{1}{1-q}$. By using
formulae (\ref{cc1}) and (\ref{c2.1})
one gets for the infinitesimal "rotations" of  matrix $g$ the following answer,
\begin{equation}
g(\epsilon )= e^{-i\epsilon H}ge^{i\epsilon H }=g-i\epsilon [H,g]+...=
	\label{d16}
\end{equation}
$$\left( \begin{array}{lr}
a+i\epsilon (1+q)cq^{N} & -q(c^{*}-i\epsilon (a^{*}q^N-q^Nae^{-2i\phi})) \\
c+i\epsilon (q^{N}a-a^{*}q^{N}e^{2i\phi}) & a^{*}-i\epsilon (1+q)q^Nc^{*}
\end{array}\right).	$$
Now consider the limit $q\to 1$. The operator of coherent states $\Psi (u)$
has the form of a sum over all spins
\begin{equation}
	\Psi (u) =1+ \Psi ^{1/2}(u) + \Psi ^{1}(u) + ...,
	\label{d17}
\end{equation}
where
\begin{equation}
	\Psi ^{j}(u) = (2j+1)\sum  W^{j *}_{mn}D^{j}_{mn}(u).
	\label{d18}
\end{equation}
We consider only the contribution from  $\Psi ^{1/2}(u) $, which is
\begin{equation}
	\Psi ^{1/2}(u) =2\mbox{tr}(ug^{*})=2(a \alpha ^{*}+ c^{*}\gamma + qc \gamma
^{*}+
a^{*}\alpha ).
	\label{d19}
\end{equation}
After a simple calculation one gets from (\ref{d16})
\begin{equation}
	<[iH,a]\Psi ^{1/2}(u)>_{q}~=-2i\frac{(1+q)(1-q^{2})}{(1-q^{5})}\gamma
{}~~\too {\longrightarrow } -i\frac{8}{5}\gamma,
	\label{d20}
\end{equation}
\begin{equation}
	<[iH,c]\Psi ^{1/2}(u)>_{q}~=-2i\frac{(1-q^{2})^{2}}{(1-q^{3})(1-q^{5})}
	\alpha
	~\too
	{\longrightarrow } -i\frac{8}{15}\alpha.
	\label{d21}
\end{equation}
Therefore for the infinitesimal rotations, one has
\begin{equation}
	<g(\epsilon )\Psi ^{1/2}(u)>_{q} ~\too
	{\longrightarrow } ~ u+i\epsilon \delta u ,
	\label{d21a}
\end{equation}
where
\begin{equation}
\delta u =\frac{4}{5}\left( \begin{array}{lr}
\gamma & \frac{1}{3}\alpha ^* \\
\frac{1}{3} \alpha &-\gamma  ^* \end{array}\right).
	\label{d22}
\end{equation}
Infinitesimal classical rotation corresponding to (\ref{d22}) is
\begin{equation}
\delta u =\left( \begin{array}{lr}
\gamma & \alpha ^* \\
\alpha &-\gamma  ^* \end{array}\right).
	\label{d23}
\end{equation}
The formulae  (\ref{d22}) and (\ref{d23})  differ by numerical
coefficients.

This consideration of quantum dynamics  is not fully
satisfactory. We would like to find such a quantum dynamics which in the
limit $q\to 1$ leads precisely
to classical dynamics.

\section{Conclusions and Discussion}

We have defined in this paper coherent states on the quantum group
$SU_{q}(2)$. One can generalize the operator of coherent states
$\Psi$ (\ref{I2}) to an arbitrary quantum group $G_{q}$ in the form
\begin{equation}
\Psi (g,u) =\sum _{\lambda}	m _{\lambda} \chi _{\lambda} (W ^{\lambda} (g)
T ^{\lambda} (u) ).
	\label{cd1}
\end{equation}
Here $\lambda $ enumerates all representations of $G_{q}$, $T^{\lambda} $
is a unitary representation of the classical Lie group $G$ , $W ^{\lambda} (g)$
is the corresponding representation of the quantum group $G_{q}$,
$\chi _{\lambda}$ is the character of the representation $\lambda$ and
$m_{\lambda }$ are numbers. It would be interesting to elaborate on such an
approach for noncompact groups.
We have noted the role of an appropriate choice of states in the
consideration of the semiclassical limit $q\to 1$ and show the existence of
classical limit for our coherent states.It would be interesting to find a
full semiclassical expansion for $<f(g)\Psi (u)>_{q}$
 like we have in standard quantum mechanics.

There are considerations of q-deformed coherent states on quantum
algebras, in particular on the q-deformation of the universal
enveloping $U_{q}(su(2))$, see
\cite{PVI}-\cite{Sol},\cite{Ell}-\cite{Demos}.
Because
of the duality between the  quantum groups and the quantum
algebra  it is possible that there is  a relation
between our  coherent states and the coherent states on quantum algebras. We
postpone a discussion of this
problem to a future publication.

The group $SU(2)$
 is a 3-dimensional manifold and there is no symplectic structure on it.
 Therefore it seems that in passing to quantum group $SU_{q}(2)$ we should
 quantize not the Poisson brackets but another invariant geometrical structure
 on $SU(2)$. There is such a structure; it is the contact structure which
 is an analog of the symplectic structure on odd dimensional manifolds
 \cite{Arn}. A contact structure on a manifold is a nondegenerate field
 of tangent hyperplanes. Such a field is given by a contact form.
If we write a contact form on $SU(2)$ locally as
 $\omega =d \phi +pdx $
then the coordinate $\phi $ is not quantized in passing to $SU_{q}(2)$
and $p$ and $x$ after complexification turn out to be the operators
$b$ and $b^{*}$ in
(\ref{r}). Note, parantethically, that in fact in gauge theory one also
deals with contact geometry since due to the constraints one has four
coordinates $A_{\mu}$
and only three canonical momenta $p_{i}$ and a natural contact form is
$\omega =d A_{0}+p_{i}dA_{i}$.

Note that quantum groups from the point of view of geometric quantization
are considered in \cite{Aru}. Deformations of Poisson brackets
on Lie groups and quantum duality are discussed in \cite{Sem}.

We have considered the semiclassical limit  $q\to 1$ by using coherent
states. In standard quantum mechanics there is an approach to
quantization involving  deformation theory using Wigner distribution and the
Moyal brackets \cite{Flato}. In such an approach one can do semiclassical
limit
directly for operators. It would be very interesting to elaborate on  an
operator expansion corresponding to our coherent states.

$$~$$
{\bf ACKNOWLEDGEMENTS}
$$~$$
 This work has been supported in part by an operating grant from the
 Natural Sciences and Engineering Research Council of Canada. I.Ya.A., R.P.
 and I.V.V. thank the Department of Physics for kind hospitality during
 their stay at Simon Fraser University.

$$~$$


\begin{thebibliography}{99}
{
\bibitem{AV}  I.Ya. Aref'eva and I.V.  Volovich, Mod.Phys.Lett., A6 (1991) 893.
\bibitem{IP1} A.P. Isaev and Z. Popovicz, Phys. Lett. 281 (1992) 271.
\bibitem{AB} V.P.Akulov, V.D.Gershun and A.I.Gumenchuk, Pis'ma
Zh.Eksp.Theor.Fiz. 56 (1992)177
\bibitem{Wu} K. Wu and R.J. Zhang, Comm. Theor. Phys.  17 (1992) 175-182.
\bibitem{H} M. Hiroyama, Prog. Theor.  Phys. 88 No 1 (1992).
\bibitem{W} S.Watamura, {\it Bicovariant Differential Calculus and
q-deformation of Gauge Theory,} preprint HD-THEP-92-45.
\bibitem{C} L. Castellani, Phys. Lett. B292 (1992) 93; \\
P.Aschieri and  L. Castellani, preprint CERN-TH 6621/92.
\bibitem{ArA} I.Ya. Aref'eva and G.E. Arutyunov,
 J. Geometry and Physics, 11(1993).
\bibitem{Maj}
T. Brzezinski and S. Majid, Phys. Lett. B 298 (1993) 339.
\bibitem{IP2}
A.P. Isaev and Z. Popovicz, {\it Quantum Group Gauge Theories and Covariant
Quantum Algebras}, Dubna, Prep. JINR E2-93 (1993).
\bibitem{AAr} I.Ya. Aref'eva and G.E.  Arutyunov, {\it Uniqueness of
$U_{q}(N)$ as a quantum gauge group  and representations of its
differential algebra,} preprint SMI-8-93.
\bibitem{Dr} V.G. Drinfeld,in Proc. ICM Berkeley CA
(Providence,RI:AMS) ed. A.M. Gleason, 798 (1986).
\bibitem{Jim} M. Jimbo, Lett. Math. Phys.,11 (1986) 242.
\bibitem{Wor} S. Woronowicz, Comm. Math. Phys. 111 (1986) 613;
Publ.RIMS Kyoto Univ. 23 (1986) 117.
\bibitem{Fad}  L.D. Faddeev, N. Reshetikhin and L.A. Tachtadjan, Alg. Anal.
1 (1988) 129.
\bibitem{AVI}
I.Ya. Aref'eva and I.V.  Volovich, Phys. Lett.  B264 (1991) 62.
\bibitem{Luk} Y. Frishman, J. Lukierski, W.J. Zakrzewski,
{\it Quantum Group $\sigma$ Model}, DTR-19/92 (1992).
\bibitem{AVII}
I.Ya. Aref'eva and I.V.  Volovich, Phys.Lett. B 268(1991)179 .
\bibitem{Rim} J. Rembieli\'nski, Phys. Lett. B287(1992)145;
in {\it From the Rotation Group to Quantum Algebras} , ed.B.Gruber
(Plenum, Nwe York) 1993; \\ T. Brzezinski, J. Rembielinski, K.A.
Smolinski, Mod.Phys.Lett. A8 (1993) 409;\\ J. Rembieli\'nski
and W.Tybor,
preprint, KFT UL 5/93, hep-th/9308058.
\bibitem{Ubr} M. Ubriaco, Phys. Lett.A 163(1992) 1; {\it Quantum Group
Schr\"{o}dinger Field Theory}, preprint LTP-035-UPR.
\bibitem{Wess} J. Schwenk and J. Wess, Phys. Lett. B291 (1992) 273.
\bibitem{Koz} S.V. Kozyrev, Theor. Math. Phys. 93 (1992) 87.
\bibitem{MM} V.I. Man'ko and R.V. Mendes, {\it q-deformed Brownian Motion},
preprint, CERN-TH-6838/93.
\bibitem{Manko} V.I.Man'ko, G.Marmo,S.Solimeno and F.Zaccaria, INFN preprint,
DSF-T-92/25.
\bibitem{Mad} S. Majid, Int. J. Modern Physics A 5(1) (1990) 1-91.
\bibitem{Kl} J.R. Klauder, {\it A coherent State Primer}: in Coherent
States, (World
Sci,1985) eds: J.R. Klauder and B.S. Skagerstam.
\bibitem{Coh} A.M. Perelomov, {\it Generalized Coherent states and their
applications}  (Springer, Berlin) , (1986).
\bibitem{PVI} R. Parthasarathy and K.S. Viswanathan, J. Phys. A: Math. Gen. 24
(1991) 613.
\bibitem{PVII} R. Parthasarathy and K.S.Viswanathan, J.Phys. A: Math.Gen. 25
(1992) L 335.
\bibitem{RefDem} E. Celeghini, M. Rasetti and G. Vitiello, Phys. Rev. Lett.
66(1991) 2056.
\bibitem{Sol} J. Katriel and A. Solomon, J. Phys.: Math. Gen. 24(1991) 2093
\bibitem{LQGG} I.Ya. Aref'eva,  G.E.Arutyunov, K.S. Viswanathan and
I.V.  Volovich, {\it Hot Lie Groups and q-deformed Itzykson-Zuber
integral}, preprint SFU-HEP-111-93.
\bibitem{Zachos} C. Zachos, {\it Thermodynamic q-Distributions That
Aren't}, preprint, UW/PT-93-05, Hep-th/ 9305067.
\bibitem{Bied}  L.C. Biedenharn, J. Phys. A, Math. Gen. 22 (1989) L873.
\bibitem{Mac}  A.J. Macfarlane, J. Phys.A, Math. Gen.22 (1989) 4581.
\bibitem{Ell} M. Chaichian, D.Ellinas and P. Presnajder, Jour. Math. Phys. 32
(1991) 3381
\bibitem{Baul} L. Baulieu and E.G. Floratos, Phys.Lett. 258 B (1991) 171
\bibitem{Demos} D. Ellinas, {\it Path integral for Quantum Algebras and the
Classical limit }, FTUV /93-7.
\bibitem{LS} L.L. Vaksman and  Ya.S. Soibelman,
Functional Anal.Appl.22(1988)170.
\bibitem{Jap} T. Masuda, K. Mimachi, Y. Nakagami, M. Noumi and K. Ueno,
J. Funct. Anal. 99 (1991) 357.
\bibitem{Koor} T.H. Koornwinder, {\it Orthogonal polynomials in connection
with quantum groups } 257 in " Orthogonal polynomials: Theory and practice.
(ed P.Nevai), NATO ASI series, C, vol.294, Kluwer, 1990.
\bibitem{Ko} H.T. Koelink, {\it On
Quantum Groups and $q$-special Functions}, PhD.  thesis, Amsterdam,1991.
\bibitem{Arn} V.I. Arnold, Mathematical Methods of Classical Mechanics,
Springer Verlag, New York, 1978.
\bibitem{Flato} F. Bayen, M. Flato, C. Fronsdal, A. Lichnerowicz and D.
Sternheimer, Ann. Phys. 111 (1978) 61.
\bibitem{Aru} G.E.Arutyunov, {\it Representations of the compact quantum
group $SU_q (N)$ and geometrical quantization}, Preprint SMI-6-93
\bibitem{Sem} M.A.Semenov-Tian-Shansky, {\it Poisson Lie Groups,
Quantum Duality Principle and the Quantum Double}, Preprint, 1993}
\end{thebibliography}
\end{document}